\pdfoutput=1
\documentclass{article}

\usepackage{arxiv}

\usepackage[utf8]{inputenc} 
\usepackage[T1]{fontenc}    
\usepackage{hyperref}       
\usepackage{url}            
\usepackage{booktabs}       
\usepackage{amsfonts}       
\usepackage{nicefrac}       
\usepackage{microtype}      
\usepackage{lipsum}
\usepackage{graphicx}
\graphicspath{ {.} }
\usepackage{doi}
\usepackage{authblk}
\usepackage{wrapfig}
\usepackage{threeparttable}
\usepackage{caption}

\title{A Disease-Specific Foundation Model Using Over 100K Fundus Images: Release and Validation for Abnormality and Multi-Disease Classification on Downstream Tasks}

\author[1,2]{Boa Jang}
\author[2,3]{Youngbin Ahn}
\author[4]{Eun Kyung Choe}
\author[5]{Chang Ki Yoon}
\author[5,6]{Hyuk Jin Choi$^{*}$}
\author[2,7]{Young-Gon Kim\thanks{Equally contributed to this work and share corresponding authorship.}$^{*}$}

\affil[1]{Interdisciplinary Program in Bioengineering, College of Engineering, Seoul National University}
\affil[2]{Department of Transdisciplinary Medicine, Seoul National University Hospital}
\affil[3]{Interdisciplinary Program of Medical Informatics, Seoul National University College of Medicine}
\affil[4]{Department of Surgery, Seoul National University Hospital Healthcare System Gangnam Center}
\affil[5]{Department of Ophthalmology, Seoul National University College of Medicine}
\affil[6]{Department of Ophthalmology, Seoul National University Hospital Healthcare System Gangnam Center}
\affil[7]{Department of Medicine, Seoul National University College of Medicine}

\renewcommand{\shorttitle}{Fundus}

\begin{document}
\maketitle

\begin{abstract}
Artificial intelligence applied to retinal images offers significant potential for recognizing signs and symptoms of retinal conditions and expediting the diagnosis of eye diseases and systemic disorders. However, developing generalized artificial intelligence models for medical data often requires a large number of labeled images representing various disease signs, and most models are typically task-specific, focusing on major retinal diseases. In this study, we developed a Fundus-Specific Pretrained Model (Image+Fundus), a supervised artificial intelligence model trained to detect abnormalities in fundus images. A total of 57,803 images were used to develop this pretrained model, which achieved superior performance across various downstream tasks, indicating that our proposed model outperforms other general methods. Our Image+Fundus model offers a generalized approach to improve model performance while reducing the number of labeled datasets required. Additionally, it provides more disease-specific insights into fundus images, with visualizations generated by our model. These disease-specific foundation models are invaluable in enhancing the performance and efficiency of deep learning models in the field of fundus imaging.
\end{abstract}

\keywords{Fundus Images \and Disease-Specific Foundation Model \and Transfer Learning \and Medical Imagining \and Dataset Efficiency}

\section{Introduction}

Vision plays a crucial role in determining quality of life, and its significance becomes increasingly pronounced with advancing age. Prevalent eye diseases like age-related macular degeneration (AMD), glaucoma, diabetic retinopathy (DR), retinal vein occlusion (RVO), pathologic myopia (PM), and epiretinal membrane (ERM) significantly affect many people and linked to the blindness \cite{campochiaro2015molecular}. Early diagnosis and prevention of these diseases are critical, particularly in light of rising prevalence rates for AMD, glaucoma, and DR, which are major eye diseases that cause blindness \cite{yau2012global, 3}. However, a recent study forecasted a sizeable shortage of ophthalmology workforce supply relative to demand by the year 2035 \cite{4}.Therefore, it is required to develop efficient screening and diagnosis system which can be easily accessed and used by not only ophthalmologists but also general physicians and even public. 

Advancements in fundus imaging technologies have led to the development of artificial intelligence (AI) screening systems, which are user-friendly, resource efficiency, and suitability for implementation in primary healthcare settings\cite{5}.These systems hold the potential to facilitate early detection of fundus abnormalities, providing critical treatment advice or referrals. Their deployment, especially in primary care areas, is anticipated to become a prevailing trend. Nevertheless, the development of highly accurate AI models for healthcare faces significant hurdles, including stringent personal data regulations and the high costs associated with data annotation\cite{6}.

\begin{figure}[t]
    \centering
    \includegraphics[width=0.8\textwidth]{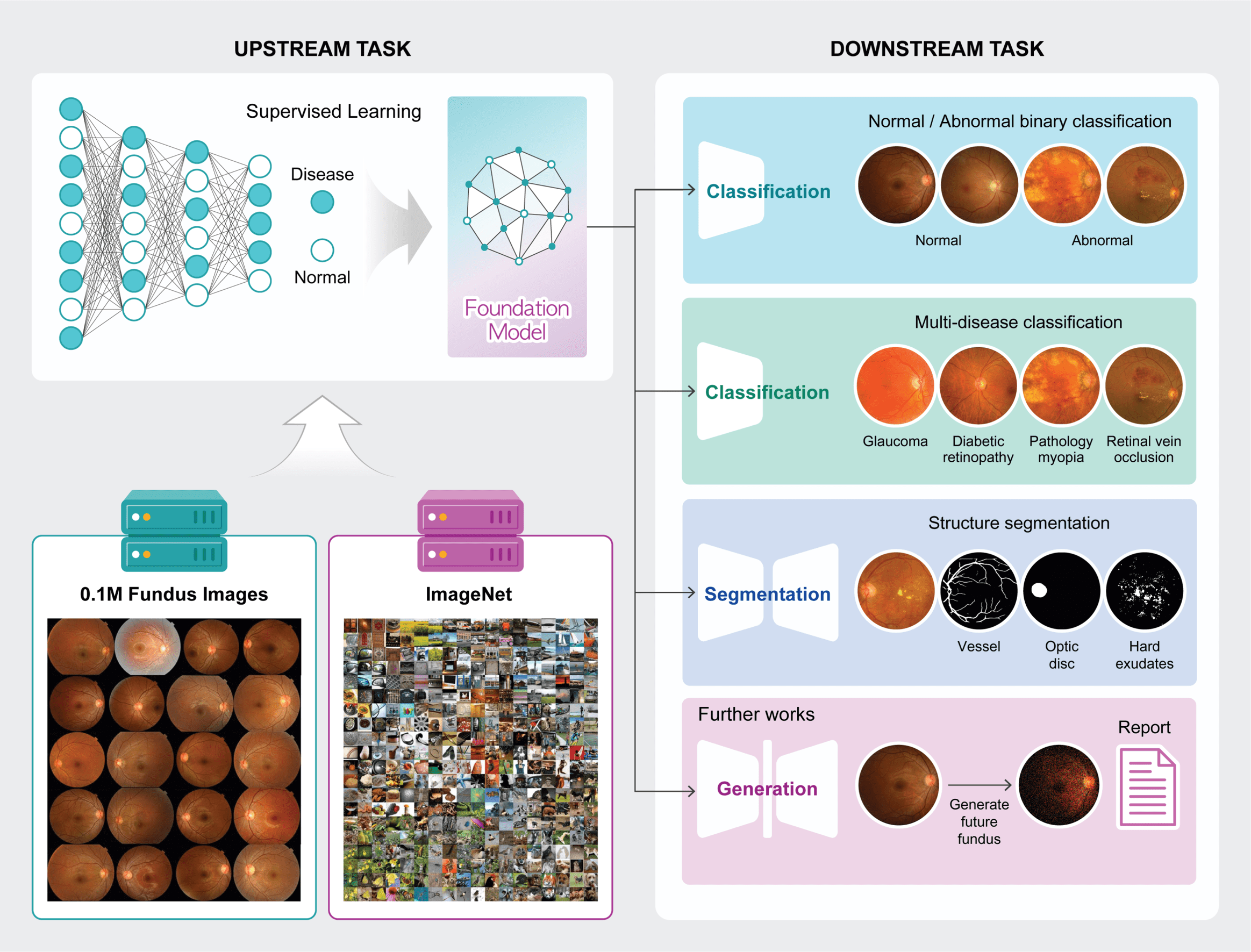}
    \caption{Illustrates the application of a pretrained fundus-specific model in processing and enhancing fundus images.}
    \label{figure_1}
\end{figure}

Transfer learning emerges as a powerful strategy in this context, leveraging knowledge from pre-trained models to enhance the effectiveness and efficiency of deep learning models in medical applications\cite{7}. Although ImageNet pre-trained models are commonly used in medical imaging tasks, their performance and applicability are often limited due to the mismatch in image resolution, as medical images typically have much higher resolutions than the 224×224 pixels used for ImageNet models \cite{8}.

In response to these challenges, domain-specific pre-trained models have shown promise. For instance, in Chest X-ray images, pre-trained models developed from extensive X-ray datasets have outperformed those based on ImageNet pre-trained weights\cite{9}. Studies have also indicated that a three-step transfer learning approach can outperform the conventional transfer learning method when applied to X-ray datasets\cite{10}. Despite the availability of a foundation model for fundus images\cite{11}, these pre-trained methods and models have not been widely accessible to researchers. 

This study introduces two novel types of disease-specific foundation models specifically designed for fundus images, developed using an extensive fundus image dataset, as shown in Figure \ref{figure_1}. These disease-specific foundation models are intended to be made freely available to the research community, addressing the gap in domain-specific tools for comprehensive disease detection and diagnosis in fundus imagery.

\section{Materials and Methods}

\subsection{Dataset of Fundus Images}

Transfer learning involves two main components: the upstream task and the downstream task. The upstream task focuses on training a model using a large dataset to generate pre-trained weights. These pre-trained weights are then utilized in the downstream task, which employs a smaller dataset to assess the efficiency and performance of the pre-trained weights.

For the upstream task in our study, fundus images were acquired retrospectively from Seoul National University Hospital Healthcare System Gangnam Center in South Korea. A total of 113,645 fundus images of 57,803 patients collected between 2003 and 2010 were used. These fundus images were labeled with normal and abnormal, and the ratio of normal and abnormal data was 8:2.

For the downstream tasks, fundus images with confirmed abnormalities were collected independently from the same hospital where the upstream data was collected. The dataset consisted of 18,459 images from 9,419 patients for the downstream tasks. Among them, 2,559 fundus images were labeled as abnormal and further classified into seven distinct classes representing different eye diseases for multi-label classification. These classes included AMD, glaucoma, glaucoma suspect, DR feature, PM, ERM, and RVO. Abnormal images which did not belong to above seven classifications were designated as “other”. This dataset exhibited significant class imbalance, with most samples belonging to the major diseases. Furthermore, certain images were excluded from the analysis due to issues such as blur and defocus.

Three public datasets were used for external validation and to generalize the tasks: RFMiD\cite{12}, JSIEC \cite{13}, and FIVES \cite{14} dataset. From the RFMiD dataset, a total of 1,920 fundus images were used for validation, labeled with conditions such as normal, AMD, glaucoma, DR, and PM. The JSIEC datasets comprised 1,000 fundus images across 39 classes and a subset of 484 cases was used for validation, labeled with conditions including normal, glaucoma, DR, and PM. The FIVES dataset, used for the vessel segmentation task, consisted of 800 fundus images each accompanied by a vessel segmented mask.

\subsection{Training Visual Representation of Fundus as an Upstream Method}

Two types of disease-specific foundation models were developed to enhance the final performance of predictive tasks. One of the models was used a large fundus image dataset with derived labels (abnormal and normal) to establish a disease-specific foundation model, referred to as ‘Fundus’. The other one includes two-steps of training, which starts with training on a large nonmedical dataset (ImageNet-1k) and was subsequently retrained using the large fundus image dataset, referred as ‘ImageNet + Fundus’. This two-step pre-training approach is designed to capture both general and medical-specific visual representations. The training employed supervised learning methods to classify the normal and abnormal of fundus image. To optimize the model for fundus image analysis, three different image resolutions were tested: 256, 512, and 1024 pixels.

The models were trained using 4 NVIDIA A100 GPUs with a batch size of 32. All implementations were carried out in PyTorch, with ImageNet pre-trained models sourced from the TorchVision library for easy usage. In this study, a 50-layer residual network (ResNet)\cite{15}, one of most commonly used networks in deep learning was used. The Adam optimizer was configured with a learning rate of 1e-5, momentum of 0.9, and weight decay of 1e-5. Data augmentation techniques included horizontal flip, grayscale, blur, and contrast limited adaptive histogram equalization (CLAHE). Weighted cross entropy loss was used as the loss function to specifically target fundus image abnormalities. The training process spanned 100 epochs with early stopping to prevent overfitting. The total training duration was approximately five days.

\subsection{Evaluation via Various Conditions of Downstream Task}

To evaluate our disease-specific foundation model, three different downstream tasks were utilized: abnormality classification (binary classification of normal and abnormal), multi-disease classification (classification of various diseases from fundus images), and vessel segmentation (segmentation of vessels from fundus images). The quality of learned representations within the fundus-specific transfer learning model was assessed through two methods: linear probing (LP), achieved by training a linear classifier on frozen backbone weights, and full fine-tuning (FT), achieved by fine-tuning the entire model. Both LP and FT evaluations involved applying a nonlinear classifier to frozen embeddings and fine-tuning the entire model, respectively. 

The abnormality classification task was evaluated under three scenarios: (a) the effectiveness of different pre-trained models, (b) the impact of image resolutions, and (c) a stress test under data-limited conditions. In scenario (a), the binary classification task involved comparing four models: one with randomly initialized weights (referred to as ‘Scratch’), one pre-trained with ImageNet weights (referred to as ‘ImageNet’), and two pre-trained with fundus-specific weights, namely ‘Fundus’ and ‘ImageNet + Fundus’. In scenario (b), the influence of image resolution on learning representations was examined by evaluating the models at three resolutions used in the upstream task -- 256 pixels (low resolution), 512 pixels (medium resolution), and 1024 pixels (high resolution). Scenario (c) involved stress tests, where models were trained with varying data fractions (1\%, 10\%, 50\%, and additional conditions as detailed in Figure \ref{figure_2}C), testing the robustness of the disease-specific foundation models under data-limited conditions.

\subsection{Measurement and Visualization of the Embeddings}

To assess model performance, five-fold cross validation was used, with the mean and standard deviation values of each metrics calculated for the performance comparisons. The area under the receiver operating characteristic curve (AUC) was the primary metric for evaluating performance across various downstream tasks, calculated from the results of the five-fold cross-validation. Differences in AUC scores between models were tested using the area test proposed by DeLong et al.\cite{16}, which assesses whether two classifiers have statistically different AUC scores, and a p-value of less than 0.05 was considered statistically significant. 

For visualizing how embeddings differ between general models and fundus-specific pre-trained models, t-distributed stochastic neighbor embedding (t-SNE)\cite{17} was used. This method facilitates the visualization of high-dimensional data by projecting it into a lower-dimensional space. Addressing the opacity of how deep neural networks make predictions – a challenge often referred to as the “black-box” problem -- gradient-weighted class activation mapping (Grad-CAM) was used\cite{18}. Grad-CAM utilizes the gradients of any target concept flowing into the final convolutional layer to create a localization map that highlights the region most influential for predicting the target. In this study, both Grad-CAM and t-SNE was provided simultaneously to enhance the interpretability of the predicted results for each fundus image, thus aiding clinicians in applying deep learning to clinical practice.

\subsection{Code Availability}

The source code and trained models are available from \textit{\href{https://github.com/Jang-Boa/Research-Foundation-Retina}{https://github.com/Research-Foundation-Retina}}.The code for implementing the two proposed disease-specific foundation models are also available for download and use. These resources aim to facilitate replication of our results and further research in the field.

\section{Results}
    
\subsection{Abnormality Classification Task} 

To address the genuine data distribution and varied conditions in real clinical environments, we considered various data and experimental settings. The Jaccard index was employed to determine thresholds.

\begin{table}[t]
 \caption{Abnormality classification performance comparison of four model types across different image sizes.}
  \centering
  \begin{threeparttable}
  \renewcommand{\arraystretch}{1.2}
  \resizebox{\textwidth}{!}{
  \begin{tabular}{lcccccc}
    \hline
    \textbf{AUC}&\textbf{256}&\textbf{\textit{p}-value}&\textbf{512}&\textbf{\textit{p}-value}&\textbf{1024}&\textbf{\textit{p}-value}\\ \hline\hline
    \textbf{Scratch}&0.698 ± 0.008&< 0.001&0.731 ± 0.008&< 0.001&0.740 ± 0.015&< 0.001\\
    \textbf{ImageNet (LP)}&0.772 ± 0.002&< 0.001&0.767 ± 0.007&< 0.001&0.734 ± 0.006&< 0.001\\
    \textbf{Fundus (LP)}&0.777 ± 0.002&< 0.001&	0.888 ± 0.003&< 0.001&0.913 ± 0.002&< 0.001\\
    \textbf{ImageNet + Fundus (LP)}&0.923 ± 0.006&	\textit{Ref.}&0.945 ± 0.002&	\textit{Ref.}&0.945 ± 0.001	&\textit{Ref.}\\
    \textbf{ImageNet (FT)}&0.877 ± 0.007&	< 0.001&	0.897 ± 0.006&	< 0.001&0.907 ± 0.003&	< 0.001\\
    \textbf{Fundus (FT)}&0.767 ± 0.009&	< 0.001&	0.877 ± 0.007&	< 0.001&0.907 ± 0.006&< 0.001\\
    \textbf{ImageNet + Fundus (FT)}&0.912 ± 0.006&\textit{\textit{Ref.}}&0.943 ± 0.005&\textit{Ref.}&0.938 ± 0.003&\textit{Ref.}\\
    \bottomrule
  \end{tabular}
  }
  \begin{tablenotes}       
    \item[*]\textit{p}-value was calculated by Delong’s test.
  \end{tablenotes}
  \end{threeparttable}
  \label{table_1}
\end{table}

\subsubsection{Comparison of Pre-training Weights} Table \ref{table_1} summarizes the mean AUC scores obtained through different model configurations and image resolutions under both LP and FT methods. The models trained from Scratch generally showed lower AUC scores, with a peak at 1024 pixels resolution (0.740 ± 0.015). Models pre-trained with ImageNet weights under LP method demonstrated a decline in performance as resolution increased, with AUC scores starting at 0.772 ± 0.002 at 256 pixels and decreasing to 0.734 ± 0.006 at 1024 pixels. Conversely, models pre-trained with Fundus weights under LP showed significant improvement, with AUC scores increasing dramatically from 0.777 ± 0.002 at 256 pixels to 0.913 ± 0.002 at 1024 pixels.

The combination of ImageNet + Fundus pre-training consistently yielded the highest AUC scores across all tested resolutions in the LP methos, maintaining an impressive score of 0.945 at both 512 and 1024 pixels. Similarly, under FT, the ImageNet + Fundus models demonstrated robust performance, although the highest AUC was observed at 512 pixels (0.943 ± 0.005) before slightly decreasing at 1024 pixels (0.938 ± 0.003).

\subsubsection{Impact of Image Resolution on Model Performance} 
\begin{figure}[t] 
    \centering
    \includegraphics[width=0.7\textwidth]{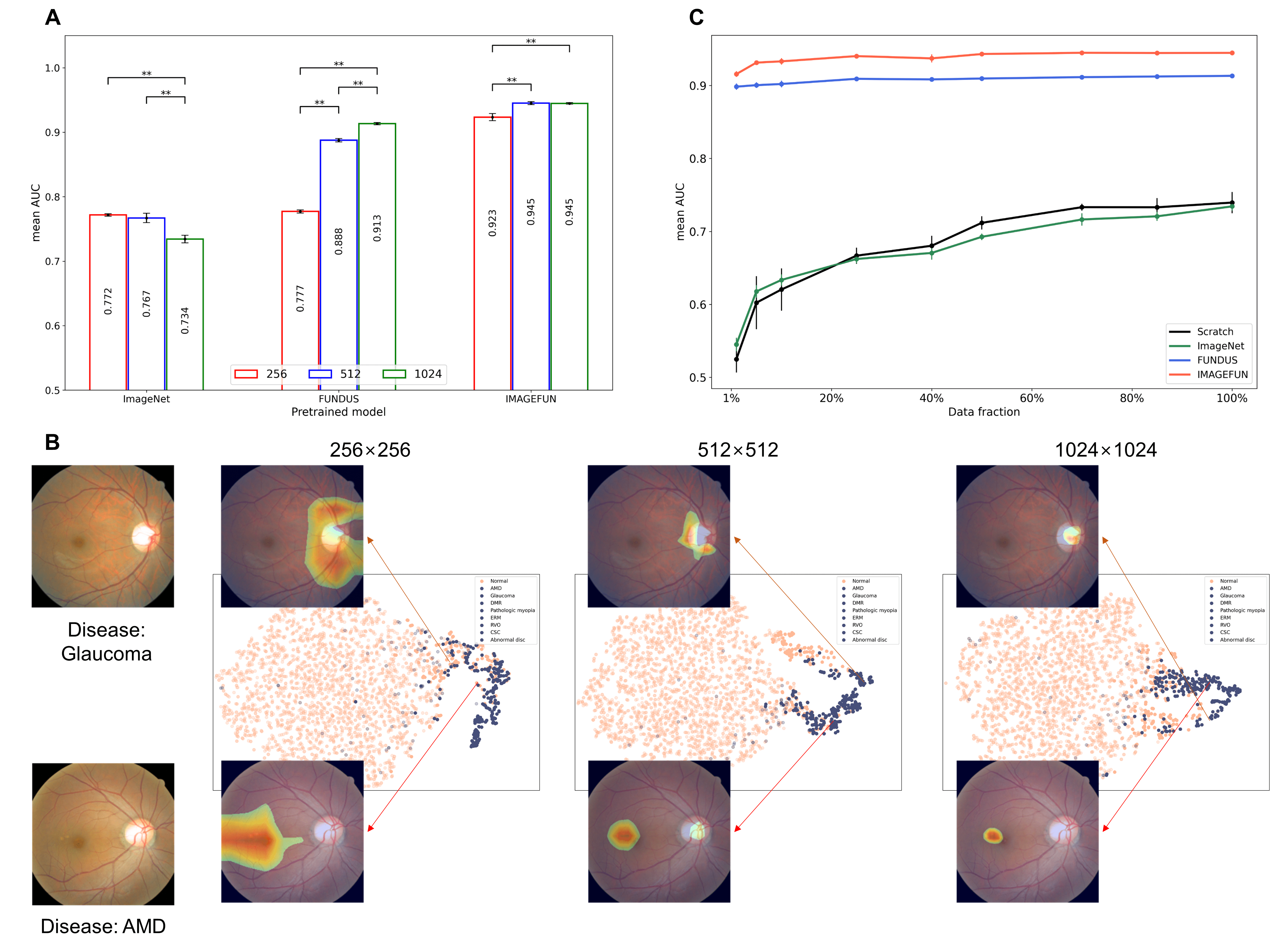}
    \caption{Comprehensive analysis of model performance across different resolutions and data fractions. (A) mean AUC scores for three pre-trained models at varying image resolution within the LF method. (B) visual representations of the activation map using Grad-CAM for the ImageNet + Fundus pre-trained model, showing areas of interest that influence the model’s predictions for the class of fundus abnormalities. (C) mean AUC scores across difference data fractions for models trained from Scratch, and those pre-trained with ImageNet, Fundus, and ImageNet + Fundus weights.}
    \label{figure_2} 
\end{figure}

Figure \ref{figure_2}A highlights the variation in mean AUC scores for different pre-trained models across increasing image resolutions within the LP method. The ImageNet pre-trained model demonstrates a decrease in performance as the resolution increases, starting with AUC of 0.772 at 256 pixels and decreasing to 0.734 at 1024 pixels. This trend suggests that the ImageNet model, originally trained on lower resolution images (224 pixels), does not adopt well to the higher resolution images typical of medical datasets.	Conversely, the Fundus pre-trained model shows a significant improvement with increasing resolution. The AUC starts at 0.777 for 256 pixels and increase to 0.913 for 1024 pixels. This indicates that the Fundus-specific training exploits the detailed information available in higher resolution fundus images more effectively, enhancing model performance substantially.

The ImageNet + Fundus pre-trained model consistently achieves the highest performance, maintaining an impressive AUC of around 0.945 for both 512 and 1024 pixels. This model benefits from a two-step pre-training approach that incorporates general visual features handling variations in image resolution.

Figure \ref{figure_2}B utilizes a t-SNE graph to compare the qualitative results of embeddings from ImageNet + Fundus pre-trained model within the LP method. This visualization provides a clear demonstration of how the model clusters and separated data points based on learned feature across different image resolutions. The activation map generated using Grad-CAM further reveal that as image resolution decreases, smaller and more specific regions are identified as critical for making clinical decisions. These findings highlight the advanced capability of the ImageNet + Fundus model to not only perform well quantitatively but also to offer insightful visual explanations that are crucial for clinical interpretation.

\subsubsection{Comparison of Stress Test} Figure \ref{figure_2}C presents the outcomes of the stress tests using various data fractions, alongside results from using the full dataset. In the full dataset condition, the ImageNet + Fundus model demonstrated superior performance, achieving the highest mean AUC across all tested conditions. This indicates that the two-step pre-training approach of combining general and specialized pre-training is highly effective for maximizing model efficacy.

 Particular attention was given to the results from the 1\% data fraction stress test, which is crucial for understanding model robustness when faced with extreme limitations in available training data. Under this stringent condition, the models trained from Scratch, with ImageNet weights, and with Fundus-specific pre-training yielded mean AUC scores of 0.522 ± 0.016, 0.545 ± 0.010, and 0.887 ± 0.003, respectively. The model combining ImageNet and Fundus pre-training achieved an even higher score of 0.910 ± 0.004. Highlighting the 1\% data fraction is particularly relevant as it underscores the resilience of the pre-trained models, especially the ImageNet + Fundus model, in scenarios closest to the worst-case real-world conditions where very little annotated data is available.

\subsection{Task Generalization}

\begin{table}[t]
 \caption{External validation results for pre-trained abnormality classification models at 1024 pixels.}
  \centering
  \renewcommand{\arraystretch}{1.2}
  \begin{threeparttable}
  \begin{tabular}{lcccc}
    \hline
    \textbf{AUC}&\textbf{JSIEC}&\textbf{\textit{p}-value}&\textbf{RFMiD}&\textbf{\textit{p}-value}\\ \hline\hline
    \textbf{Scratch}&0.810 ± 0.075&< 0.001&0.627 ± 0.063&< 0.001\\
    \textbf{ImageNet (LP)}&0.868 ± 0.040&< 0.001&0.695 ± 0.0577&< 0.001\\
    \textbf{Fundus (LP)}&0.934 ± 0.002&< 0.001&	0.855 ± 0.004&< 0.001\\
    \textbf{ImageNet + Fundus (LP)}&0.943 ± 0.007&	\textit{Ref.}&0.893 ± 0.003&\textit{Ref.}\\ \hline
    \textbf{ImageNet (FT)}&0.922 ± 0.011&	< 0.001&	0.866 ± 0.022&	< 0.001\\
    \textbf{Fundus (FT)}&0.940 ± 0.003&	< 0.001&	0.845 ± 0.010&	< 0.001\\
    \textbf{ImageNet + Fundus (FT)}&0.929 ± 0.012&\textit{\textit{Ref.}}&0.876 ± 0.008&\textit{Ref.}\\
    \bottomrule
  \end{tabular}
  \begin{tablenotes}       
    \item[*]\textit{p}-value was calculated by Delong’s test.
  \end{tablenotes}
  \end{threeparttable}
  \label{table_2}
\end{table}

\subsubsection{External Validation for Abnormality Classification} The external validation, which are crucial for assessing the generalizability and reliability of our models beyond the training environment, was conducted using two distinct datasets: JSIEC and RFMiD, and the results are summarized in Table \ref{table_2}. For the JSIEC dataset, the ImageNet + Fundus model under LP achieved the highest AUC of 0.943 ± 0.007, indicating a strong predictive performance. For the more challenging RFMiD dataset, the ImageNet + Fundus achieving AUCs 0.876 ± 0.022. These results demonstrate the model’s capability to perform effectively across different external datasets, though they also highlight the variability in performance that can arise due to differences in dataset characteristics or image quality.

\begin{table}[b]
  \caption{External validation results for pre-trained abnormality classification models at 1024 pixels.}
  \centering
  \begin{threeparttable}
  \renewcommand{\arraystretch}{1.2}
  \resizebox{\textwidth}{!}{
  \begin{tabular}{lcccccccc}
    \hline
    \textbf{F1 score}&\textbf{Threshold}&\textbf{AMD}&\textbf{Glaucoma}&\begin{tabular}{cc}
         \textbf{Glaucoma} \\
         \textbf{Suspect} 
    \end{tabular}&\textbf{DR}&\textbf{PM}&\textbf{ERM}&\textbf{RVO}\\ \hline\hline
    \textbf{Number of Disease}&&211&41&211&14&9&56&5\\ \hline
    \textbf{Scratch}&0.85&0.4629&0.1654&0.5464&0.0851&0.4103&0.1875&0.0000\\
    \textbf{ImageNet(LP}&0.85&0.5026&0.2162&0.5586&	0.0909&	0.6154&	0.3333&	0.0000\\
    \textbf{Fundus(LP)}&0.85&0.6104&0.3516&0.7865&0.1961&0.8000&0.3014&0.0000\\
    \textbf{ImageNet+Fundus(LP)}&0.80&0.8689&0.4938&0.6728&0.2667&0.8182&0.6783&0.6667\\
    \bottomrule
  \end{tabular}
  }
  \begin{tablenotes}       
    \item[*]\textit{p}-value was calculated by Delong’s test.
  \end{tablenotes}
  \end{threeparttable}
  \label{table_3}
\end{table}

\begin{figure}[t] 
    \centering
    \includegraphics[width=0.75\textwidth]{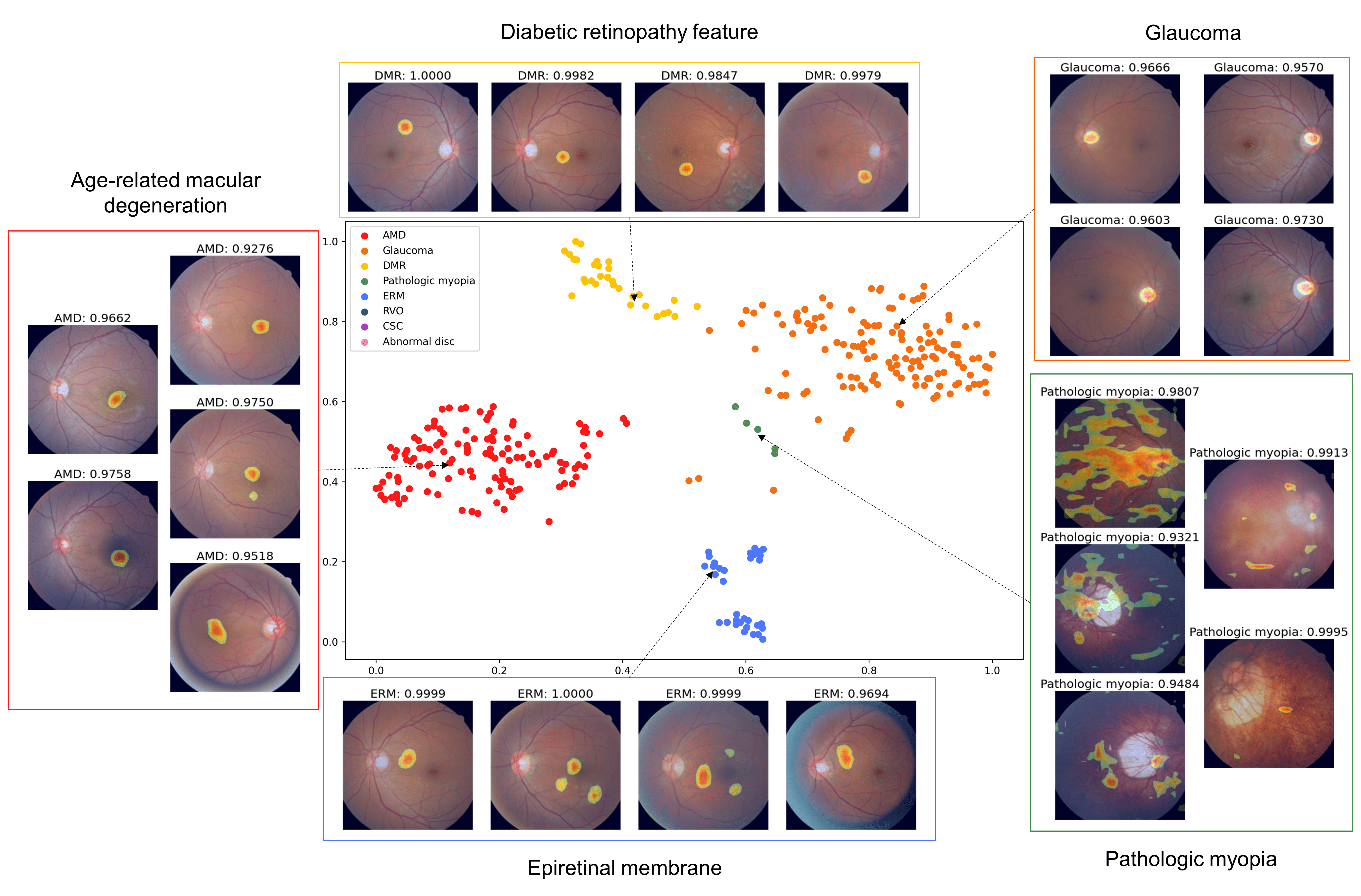}
    \caption{t-SNE visualization of multi-disease classification results with foundation model of ImageNet + Fundus.}
    \label{figure_3} 
\end{figure}

\subsubsection{Extending Model Capabilities: Multi-Disease 
Classification } To enhance the versatility of our proposed model, we extended its capabilities to multi-disease classification. Utilizing the disease-specific foundation model as a backbone, we adapted it for multi-label classification tasks by replacing the output layer to accommodate multiple class classifications for various diseases at a resolution of 1024 image size. The loss function was adjusted to cross-entropy for each label. Fine-tuning the model on a multi-label annotated dataset allowed it to learn specific disease patterns within the fundus images, enhancing its ability to accurately identify and categorize abnormal fundus images involving seven distinct disease types: AMD, Glaucoma, Glaucoma suspect, DR features, PM, ERM, and RVO.

Given the imbalanced data distribution characteristic of these disease categories, we implemented different thresholds for each model to optimize the final classification results. This approach ensures that the model's performance is robust across various types of fundus diseases, even in the presence of significant class imbalances. 

According to Table \ref{table_3}, the ImageNet + Fundus model, under the LP method, demonstrated the best performance across all disease categories. Remarkably, the ImageNet + Fundus pre-trained model showed the best overall performance across all categories, notably achieving high scores even in challenging conditions like RVO.

Further, visualizations using t-SNE revealed that the model pre-trained on fundus-specific data presented a more nuanced and detailed representation of fundus images, compared to those models trained from scratch or pre-trained on ImageNet. As shown in Figure \ref{figure_3}, the t-SNE graph for the ImageNet + Fundus model illustrates its enhanced ability to discern subtle differences between various disease signatures.

\subsubsection{Extending Model Capabilities: Vessel Segmentation} 

\begin{wrapfigure}{r}{0.58\textwidth} 
    \vspace{-1\baselineskip}
    \centering
    \includegraphics[width=0.58\textwidth]{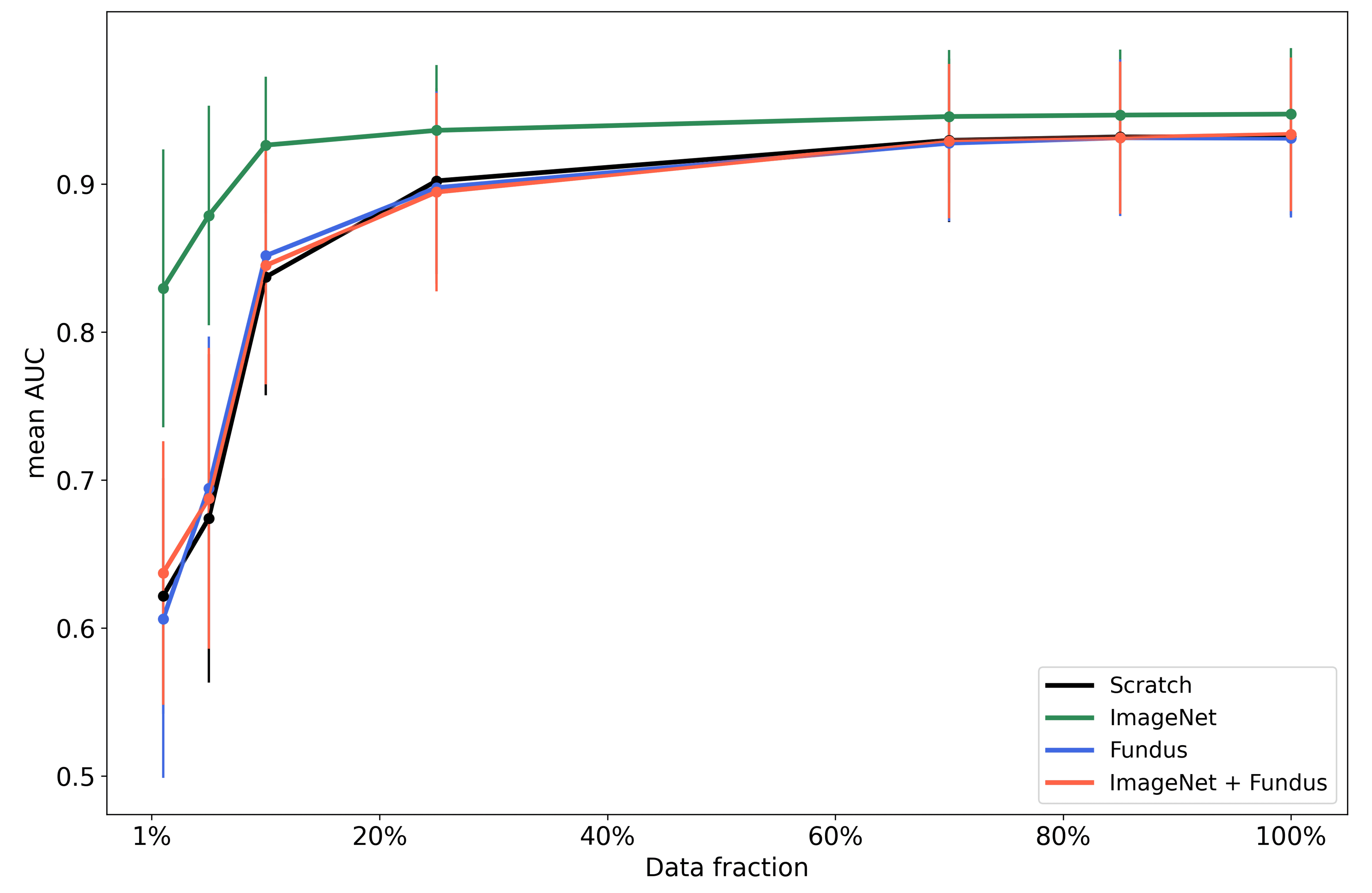}
    \caption{Performance evaluation across data fractions for vessel segmentation under the full fine-tuning method.}
    \label{figure_4} 
\end{wrapfigure}

To further demonstrate the adaptability of our disease-specific foundation model, we extended its functionality to include vessel segmentation. Using the same pre-trained model as its foundation, we developed a dedicated vessel segmentation model by incorporating a U-Net based architecture and implementing the Dice coefficient as the loss function. We employed the FIVES dataset for both training and validation, focusing on refining out model through FT methods.

The performance analysis for vessel segmentation task is detailed in Figure \ref{figure_4}, which highlighted the robust and consistent efficacy of the ImageNet-based model across various data volumes. This model's success in capturing features relevant to vessel segmentation emphasizes its applicability in tasks focused on the structural delineation within fundus images, distinguishing it as an effective tool for evaluating the specialization of our models.

\section{Discussion}
In this study, we demonstrated the superior performance of the Fundus and ImageNet + Fundus models over Scratch and ImageNet pre-trained models across a various downstream task. This superior performance was evident in both abnormality binary classification and multi-disease classification tasks, addressing challenges related to data imbalances and deficiencies effectively. 

The two-step pre-training method, particularly the ImageNet + Fundus model, consistently outperformed other models, underscoring the benefits of disease-specific foundation models. The superior performance of this model in both the JSIEC and RFMiD datasets during external validation processes indicates its robustness and generalizability across different data characteristics and qualities. The image-scale based analysis revealed that transfer learning is most effective when the data closely resembles the pre-trained model’s original training set. However, our findings suggest that models trained on ImageNet may not perform as effectively as those starting from scratch due to the mismatch between general and medical image characteristics.

In the multi-disease classification task, both of disease-specific foundation models are outperformed general pre-trained models, as it had already effectively learned all pertinent features. However, a different result emerged in the vessel segmentation task. While the models pre-trained on Fundus data were highly effective at recognizing disease patterns, they were less successful at structural tasks like vessel segmentation. This suggests that while these models are specialized, their focus on disease-related features may limit their effectiveness in tasks requiring anatomical detail recognition.

A notable finding from our study is the underperformance of the commonly utilized ImageNet pre-trained model in certain medical domains or tasks, particularly when confronted with insufficient data. This highlights the critical importance of selecting an appropriate pre-trained model that is tailored to the specific domain and task at hand. Furthermore, recognizing that most medical data consist of high-quality information, it has been established that adopting an appropriate image size is crucial. Learning high-quality data using a pre-learning model trained on low-quality images may not contribute to performance improvement. As shown in Figure \ref{figure_4}, as the image size increases, the model's heatmap can pinpoint more specific areas. While the model wasn't designed for disease classification per se, it was observed that the heatmap varied for each disease, suggesting the potential for classification based on disease-specific diagnostic characteristics. Disease diagnosis necessitates an understanding of each condition's unique features for accurate assessment. Our learned model demonstrated the ability to discern these distinctive characteristics for each disease, suggesting its utility across various fundus studies.

Based on the results of the stress tests, the disease-specific foundation models exhibit varying degrees of effectiveness in handling data shortages. The comparison across different data fractions highlights the importance of utilizing pre-trained models tailored to the specific dataset. Notably, the ImageNet + Fundus model consistently outperformed others across all conditions, even in scenarios with severely limited data (1\% fraction). This suggests that leveraging both general and domain-specific knowledge yields superior performance in overcoming data scarcity challenges. The significant performance gap between the Scratch and pre-trained models underscores the advantage of transfer learning in such contexts. Disease-specific foundation models, particularly those trained with general and Fundus images, demonstrate a remarkable ability to extract relevant features despite limited training samples. These findings emphasize the practical benefits of employing pre-trained models in real-world research environments where data availability may be constrained. By leveraging existing knowledge and adapting it to the target domain, researchers can effectively address data shortages and achieve robust performance in various tasks, including classification and prediction. Furthermore, the results underscore the need for careful consideration when selecting pre-trained models, as the choice can significantly impact performance outcomes. Future research could explore additional techniques to further optimize model performance under data scarcity conditions, potentially enhancing the applicability of deep learning approaches in real-world scenarios.

Despite its successes, the developed model faced with certain limitations. Primarily, its specialized focus on classifying fundus diseases has led to less-than-optimal performance in tasks involving the extraction of structures from fundus images beyond the scope of learned normal and abnormal information. Furthermore, severe data imbalances in disease classification have hindered performance, especially for diseases with limited data and unclear characteristics. Also, in segmentation task, our disease-specific foundation models are not outperformed. Future improvements and fine-tuning are expected to enhance model performance. Our goal is not just to develop this model and conclude the process; rather, we aim to create a foundational model that can be comprehensively utilized across various ophthalmic research endeavors. Therefore, our future research direction involves developing models capable of recognizing, extracting, and generating various structures within retina, rather than solely diagnosing specific diseases. These models will serve as versatile tools for diverse applications in ophthalmic imaging and analysis.

Our research focused on developing a disease-specific foundation model specifically tailored for fundus image analysis. Through extensive validation under diverse experimental conditions, we confirmed the superior performance of the disease-specific foundation models compared to the model which trained from scratch or pre-trained on ImageNet. This emphasizes the importance of domain-specific training and highlights the benefits in fundus image based models. Leveraging a large-scale fundus image dataset and domain-specific knowledge, our model demonstrated improved accuracy in abnormality detection and disease classification tasks.

The successful development of the disease-specific foundation model underscores its significance in medical imaging, where domain-specific features play a pivotal role. By enhancing diagnostic accuracy and facilitating early detection and management of various abnormal eye conditions, our research contributes to advancing fundus image analysis. Ultimately, this work has implications for improving patient care and healthcare practices in ophthalmology. The released model weights are anticipated to play a crucial role as an auxiliary means for swiftly distinguishing between normal and abnormal fundus diseases for ophthalmologists.

\end{document}